\documentclass[runningheads]{llncs}
\usepackage[T1]{fontenc}
\usepackage[hidelinks]{hyperref}

\usepackage{graphicx}
\usepackage{color}

\usepackage[cmex10]{amsmath}
\usepackage{color}
\usepackage{amssymb}
\usepackage{pifont}
\usepackage{multirow, hhline}
\usepackage[table]{xcolor}
\usepackage{subcaption}
\usepackage{scalerel}
\usepackage{tikz}
\usepackage{url}

\usepackage[acronym,shortcuts]{glossaries-extra}
\glssetcategoryattribute{acronym}{nohyper}{true}
\setabbreviationstyle[acronym]{long-short}

\newacronym{acric}{ACRIC}{Authenticated Cyclic Redundancy Integrity Check}
\newacronym{aic}{AIC}{Availability, Integrity, and Confidentiality}
\newacronym{bitw}{BITW}{Bump-in-the-Wire}
\newacronym{cia}{CIA}{Confidentiality, Integrity and Availability}
\newacronym{crc}{CRC}{Cyclic Redundancy Check}
\newacronym{dos}{DoS}{Denial-of-Service}
\newacronym{ds}{DS}{Digital Signature}
\newacronym{hmac}{HMAC}{Hash-based Message Authentication Code}
\newacronym{hvac}{HVAC}{Heating, Ventilation, and Air Conditioning}
\newacronym{ics}{ICS}{Industrial Control System}
\newacronym{ids}{IDS}{Intrusion Detection System}
\newacronym{iiot}{IIoT}{Industrial Internet-of-Things}
\newacronym{iv}{\texttt{InitVec}}{Initialization Vector}
\newacronym{it}{IT}{Information Technology}
\newacronym{mac}{MAC}{Message Authentication Code}
\newacronym{mitm}{MITM}{Man-In-The-Middle}
\newacronym{ot}{OT}{Operating Technology}
\newacronym{otp}{OTP}{One-Time Pad}
\newacronym{p2p}{P2P}{Point-to-Point}
\newacronym{tls}{TLS}{Transport Layer Security}

\begin{document}
\title{ACRIC: Securing Legacy Industrial Networks via Authenticated Cyclic Redundancy Integrity Check}
\titlerunning{Authenticated Cyclic Redundancy Integrity Check}  

%
%

\author{
Alessandro Lotto\orcidID{0000-0003-3556-4589}\and
Alessandro Brighente\orcidID{0000-0001-6138-2995}\and
Mauro Conti\orcidID{0000-0002-3612-1934}}

\authorrunning{A. Lotto et al.}
%
\institute{University of Padua, Department of Mathematics, Italy\\
alessandro.lotto@math.unipd.it\\
\{alessandro.brighente, mauro.conti\}.@unipd.it}
\maketitle              
\begin{abstract}
The increasing integration of modern Information Technology (IT) into Operational Technology (OT) and industrial systems is expanding the vulnerability surface of legacy infrastructures, which often rely on outdated protocols and resource-constrained devices.
Recent security incidents in safety-critical industries exposed how the lack of proper message authentication enables attackers to inject malicious commands or alter system behavior, revealing fundamental security weaknesses in existing architectures.
These shortcomings have thus prompted new regulations that emphasize the pressing need to strengthen cybersecurity, particularly in legacy systems.
Message authentication is widely recognized as a fundamental security measure that enhances system resilience.
However, its adoption in legacy industrial environments is limited due to practical challenges like backward compatibility, message format changes, and hardware replacement or upgrades costs.

In this paper, we introduce ACRIC, a message authentication solution to secure legacy industrial communications explicitly tailored to overcome those challenges all at once.
ACRIC uniquely leverages cryptographic computations applied to the Cyclic Redundancy Check (CRC) field - already present in most industrial communication protocols - ensuring robust message integrity protection and authentication without requiring additional hardware or modifications to existing message formats.
ACRIC's backward compatibility and protocol-agnostic nature enable coexistence with non-secured devices, thus facilitating gradual security upgrades in legacy infrastructures.
Formal security assessment and experimental evaluation on an industrial-grade testbed demonstrate that ACRIC provides robust security guarantees with minimal computational overhead (\(\sim 4 \mu s\)).
These results underscore ACRIC's practicality, cost-effectiveness, and suitability for effective adoption in resource-constrained industrial environments.

\end{abstract}
%
%


\section{Introduction}  \label{sec:Introduction}

Industries such as power generation, oil and gas facilities, water treatment plants, and nuclear installations, which operate critical infrastructures through \ac{ot} and \ac{ics} are increasingly prioritizing cybersecurity to protect their systems from unauthorized access and potential harm~\cite{CybersecFor4.0}.
Legacy systems, characterized by outdated protocols and resource-constrained hardware, are typically prevalent on these infrastructures.
Recent high-profile incidents, including the 2010 Stuxnet attack on Iranian nuclear centrifuges~\cite{Stuxnet}, the 2015 Ukraine power grid breach~\cite{UkraininianAttack}, and the 2017 LogicLocker attack on a U.S. water treatment plant~\cite{LogicLocker}, demonstrated the severe implications of inadequate security measures on national and public safety.
This urgency in strengthening cybersecurity posture of legacy and safety-critial infrastructures is further reinforced by regulatory efforts like the forthcoming European Cyber-Resilience Act~\cite{Cyber_act} and the 2021 U.S. Executive Order~\cite{US_order}.
However, implementing modern cybersecurity in \ac{ot} and \acp{ics} poses practical challenges.
Such systems inherently lack essential security features such as authentication, integrity protection, and confidentiality, and their critical role often necessitates uninterrupted availability, complicating upgrades and introduction of advanced security solutions \cite{IndustrialPerspective}, \cite{SecurityChallenges}.
Furthermore, while beneficial, the growing convergence with \ac{it} removes the historical assumption of system isolation and trustworthiness, thereby expanding the vulnerability surfaces and exposure to new and modern cyber threats \cite{CybersecFor4.0}, \cite{IoTsecurity}.
Nonetheless, this security dilemma extends beyond industrial infrastructure, affecting other cyber-physical systems such as in-vehicle networks, as exemplified by the 2018 Jeep hacking demonstration~\cite{CANExploit}.

Although retrofitting security into legacy protocols is widely recognized as necessary, proposed solutions in the literature have not seen extensive industrial adoption \cite{ICSSurvey}.
This limited uptake is primarily due to their narrow protocol-specific focus, insufficient consideration of practical deployment constraints such as interoperability, data throughput, computational overhead, and the costs associated with additional hardware deployment \cite{InvehicleNetworks}, \cite{RetrofittingProtection}.
This highlights the need for an authentication solution adaptable across multiple systems that effectively addresses these real-world constraints.
In response, this paper presents \textbf{\ac{acric}}, an authentication framework explicitly developed to secure communications within legacy protocols.
\ac{acric} enhances security by cryptographically computing the \ac{crc} field \cite{CRC} — already present in most industrial communication protocols — leveraging a secret initialization vector and \ac{otp} encryption.
\ac{acric} introduces authentication and integrity without modifying existing message formats, adding extra fields, or requiring additional hardware - challenges not simultaneously addressed by proposed solutions.
\ac{acric}'s backward compatibility and maintained data throughput facilitate interoperability between secure and non-secure devices, enabling a gradual transition toward enhanced security.
Moreover, the leverage of the already existing \ac{crc} field makes \ac{acric} protocol agnostic, thus ensuring it is (\textit{i}) compatible with both \ac{p2p} and broadcast communications, (\textit{ii}) applicable across diverse system types, and (\textit{iii}) cost-effective for deployment.
Experimental testing with commercially available devices shows that \ac{acric} incurs minimal overhead (\(\sim 4 \mu s\)), preserving real-time system capabilities and outperforming common \ac{hmac} solutions.

\indent \ac{acric} is particularly advantageous for legacy networks, where new hardware deployment is challenging, and continuous availability is essential.
Leveraging existing \ac{crc} fields for authentication inherently limits security to the finite length of the \ac{crc}, which may be vulnerable to brute-force attacks.
This limitation can be effectively managed, for example, by setting a threshold for rejected messages to trigger security alerts - a common practice in the \ac{ics} domain \cite{CAN}, \cite{modbus-application}.
While modern methods like Digital Signatures or longer \acp{mac} might suit new systems without stringent constraints better, \ac{acric} remains optimal for securing legacy systems.\\
\textbf{Contributions.} In summary, our contributions are as follows:
\begin{enumerate}
    \item We identify critical challenges in designing security solutions for legacy industrial protocols that currently limit the adoption of state-of-the-art solutions.

    \item We develop \ac{acric}, an authentication mechanism that offers message authentication and integrity protection, explicitly tailored to legacy environments and suitable for \ac{p2p} and broadcast communications.
    In addition, we also provide practical insight for the efficient management of cryptographic material in legacy contexts. 

    \item We formally evaluate \ac{acric}'s security guarantees and experimentally evaluate its transmission and computational overhead in a realistic test environment, demonstrating minimal latency impact (\(\sim 4 \mu s\) overhead) suitable for practical deployment.
\end{enumerate}

The paper is organized as follows.
Section~\ref{sec:Related} reviews the related works.
Section~\ref{sec:Challenges} outlines the identified challenges when integrating an authentication mechanism in legacy systems, and the corresponding \ac{acric}'s design goals to address them.
Section~\ref{sec:Model} defines the system and threat models.
Section~\ref{sec:ACRIC} details the proposed framework and Section~\ref{sec:Evaluation} carries out security and performance evaluation.
Finally, Section~\ref{sec:Conclusions} concludes the paper.
\section{Related Works}\label{sec:Related}
Designing authentication mechanisms for legacy communication protocols in resource-constrained industrial environments presents significant challenges.\\
Legacy systems typically feature limited computational capabilities, restricted memory, and predefined protocols lacking built-in security.
Introducing robust authentication without affecting system performance or interoperability is therefore challenging.
Several proposals rely on asymmetric encryption, digital signatures, digital certificates, or the integration of secure protocols like \ac{tls}~\cite{Ferst-et-al}, \cite{authentiCAN}, \cite{modbus-security}, \cite{ADSB-1}, \cite{Shahzad-et-al}, \cite{ADSB-2}.
Although these methods offer strong security, they require computationally intensive cryptographic operations.
As such, they introduce significant latency, memory usage, and computational overhead, making them impractical for real-time and resource-constrained systems.
Conversely, \ac{acric} employs lightweight \ac{crc} and \ac{otp} operations, significantly reducing resource usage and overhead.\\
\indent
To mitigate these constraints, many researchers adopted \acp{mac}, particularly \acp{hmac} \cite{lightweightMAC}, \cite{Libracan}, \cite{Liu-et-al}, \cite{vatiCAN}, \cite{Pricop-et-al}, \cite{Stancu-et-al}, \cite{VulCAN}, \cite{Xuan-et-al}.
Hash functions are more efficient than asymmetric algorithms and suitable for constrained devices.
However, (H)\acp{mac} are typically appended to or integrated within the payload, requiring additional security parameters like counters, nonces, or timestamps.
This modifies standard message frames, reducing data throughput, performance, and efficiency.
Additionally, altering frame structures negatively impacts backward compatibility and interoperability with non-secured devices.
In contrast, \ac{acric} maintains the original frame structure without adding extra fields or parameters, preserving throughput and compatibility.\\
\indent
Another common strategy involves a \ac{bitw} approach, offloading cryptographic tasks to external devices placed between legacy systems and the network~\cite{Fovino_et_al},~\cite{Katulic-et-al}, \cite{Rajesh-et-al}, \cite{Tidrea-et-al}, \cite{Tsang-et-al}, \cite{Wright-et-al}.
Although this approach reduces computational demands on legacy hardware, it introduces additional components, increasing complexity, costs, and potential single-point of failure.
\ac{acric} avoids additional hardware, relying solely on lightweight cryptographic operations, preserving co-existence between secure and non-secure devices within the same network.\\
\indent
Covert channels represent another approach, embedding authentication data without altering original packet formats~\cite{Bernieri-et-al},~\cite{TACAN}.
While this maintains interoperability, it often requires specialized hardware or modifications to existing devices to embed or extract hidden information.\\
\indent
Recently, researchers explored enhancing the \ac{crc} field — originally designed for error detection — to enable authentication using secret generator polynomials for \ac{crc} computation \cite{CRC3}, \cite{CRC7}, \cite{CRC2}, \cite{CRC4}, \cite{CRC5}.
However, selecting suitable polynomials that balance cryptographic strength with error detection capability is computationally demanding and not suitable for devices with limited resources~\cite{CRCC-hard}.
\ac{acric} overcomes this limitation by using a secret \ac{iv} rather than changing the polynomial, and secures authentication by encrypting the \ac{crc} with \ac{otp}.
This avoids exhaustive polynomial searches and retains standard error detection capabilities.
Hence, \ac{acric} provides robust security without high computational demands.\\
\indent
In summary, \ac{acric} effectively addresses all practical challenges involved in securing legacy communication protocols at once - an achievement that other proposals either fail to accomplish or only partially address.
It offers a lightweight, secure, and compatible authentication mechanism, preserving system performance, data throughput, and interoperability without requiring hardware changes.
\section{Securing Legacy Systems: Design Challenges and Goals}\label{sec:Challenges}

\begin{figure}[!h]
    \centering
    \includegraphics[scale=0.5]{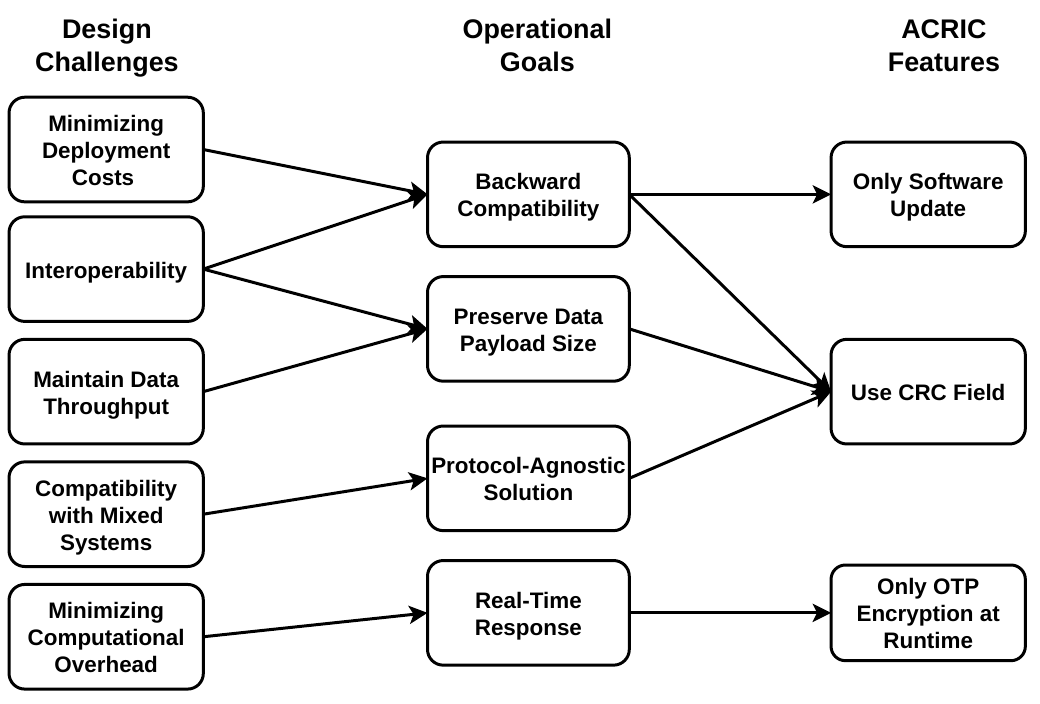}
    \caption{Relationship between identified design challenges for authentication mechanisms in legacy systems, derived operational goals to address them, and corresponding \ac{acric} key features to achieve such goals.}
    \label{fig:Layout}
\end{figure}

This section discusses the primary challenges encountered when designing authentication mechanisms for legacy industrial communication protocols, which significantly limit the real-world adoption of proposed solutions in the literature.
To identify key constraints and practical challenges, we conducted a comprehensive review of the literature and engaged in a close collaboration with CAREL~\footnote{CAREL is a global leader in control solutions for HVAC systems with over 50 years of experience and 15 production sites worldwide. Website: \url{https://www.carel.it/}}, a multinational company specialized in \ac{ics} deployed in environments marked by stringent resource constraints and real-time operational requirements.
This collaboration provided grounded insights into real-world limitations and relevance to our analysis.
Building on these findings, we defined the specific security and operational goals that emerge from such challenges, highlighting the architectural choices that make \ac{acric} both robust and practically deployable.
Figure \ref{fig:Layout} visually represents the relationships between the identified challenges, the resulting design goals, and the corresponding features offered by \ac{acric}.

\subsection{Challenges}
Significant research aimed to improve the security of legacy communication protocols, particularly within industrial and automotive networks.
The rapid advancement of \ac{iiot} and automated driving technologies has underscored the need for secure data exchanges in previously isolated, now interconnected systems \cite{CybersecurityICS}, \cite{CybersecFor4.0}, \cite{CANSurvey}.
Although various security mechanisms such as digital signatures, \acp{mac}, and encryption have been explored (\cite{CyberattacksCountermeasures}, \cite{AnalysisSecurity}, \cite{InvehicleAuth}), practical constraints limit their adoption in the real world, especially in resources-constrained environments.
By critically reviewing the available literature, we identify the following challenges.

\textbf{\underline{Challenge 1 - Interoperability}}.
Simultaneous upgrades of all devices are impractical and costly, particularly in critical industrial settings.
Additionally, in \ac{ot} systems, devices often rely on synchronized timing and regularly monitor the communication line, expecting messages of predefined lengths.
Introducing additional fields into message formats can disrupt this synchronization, leading to communication and system failures.
Therefore, security solutions must maintain backward compatibility by preserving the original message frame structure, enabling secure and non-secure devices to coexist \cite{IndustrialPerspective}, \cite{InvehicleAuth}.

\textbf{\underline{Challenge 2 - Preserving Data Throughput}}.   
Industrial communication protocols prioritize efficient data transmission due to limited bandwidth and the need for timely data exchange.
Therefore, security measures should not reduce data throughput \cite{RetrofittingProtection}.
Including authentication data within the payload reduces available space, negatively impacting throughput.

\textbf{\underline{Challenge 3 - Minimizing Deployment Costs}}.
Several proposed solutions introduce a gateway attached to legacy devices that encrypt/decrypt or parse the received messages, making the introduced security solution transparent for the legacy device.
However, solutions requiring additional hardware or extensive infrastructure modifications are often prohibitively expensive \cite{IndustrialPerspective}, \cite{CANSurvey}.
Large-scale device replacements are typically infeasible in industrial settings, making cost-effective deployment crucial for widespread adoption of security solutions.

\textbf{\underline{Challenge 4 - Minimizing Computational Overhead}}.
Real-time performance is essential for time-sensitive systems.
Security mechanisms must be lightweight and computationally efficient to avoid compromising the real-time capabilities of these critical applications \cite{CANSurvey}, \cite{SecurityChallenges}.

\textbf{\underline{Challenge 5 - Compatibility with Mixed Systems}}.
Existing solutions target specific communication protocols, requiring different implementations for each protocol and increasing deployment costs and complexity.
An adaptable solution compatible with diverse protocols is therefore preferable \cite{RetrofittingProtection}.

\subsection{Security Goals}\label{sec:SecGoals}
Legacy industrial protocols are vulnerable to various attacks, including message spoofing, modification, replay, \ac{mitm}, and \ac{dos} \cite{Fovino_et_al}, \cite{attack_taxonomies}, \cite{SecurityChallenges}.
Due to their critical nature, unlike traditional \ac{it} systems, industrial \ac{ot} systems generally prioritize availability and data integrity over confidentiality, as data changes or disruptions can significantly impact physical operations and safety~\cite{SecurityChallenges}.
Confidentiality is typically less critical since industrial communications are often repetitive and predictable, allowing attackers to infer information without explicit decryption \cite{Castellanos-et-al}, \cite{Wright-et-al}.
Given these considerations, the primary security objectives for legacy industrial systems include \cite{AnalysisSecurity}, \cite{Fovino_et_al}, \cite{SecurityChallenges}: \textit{(i) Authentication}, thanks to which we verify the legitimacy of incoming messages and communicating nodes; \textit{(ii) Integrity Protection}, which prevents unauthorized modifications to messages preserving the accuracy and trustworthiness of the data; \textit{(iii) Replay Protection}, which prevents the reuse of previously transmitted data to avoid unauthorized actions.

\subsection{Operational Goals}\label{sec:OperationalGoals}
We present our operational goals to directly address previously identified challenges, thus ensuring an authentication mechanism to be practical and ready for real-world adoption:
\begin{itemize}
    \item \textbf{Provide Backward Compatibility} (Challenges 1 and 3).
    The solution must allow the coexistence of secure and non-secure devices without changing the original frame format or adding new hardware.
    \ac{acric} meets this goal by leveraging the existing \ac{crc} field present in most legacy protocols.

   \item \textbf{Preserve Data Payload Size} (Challenge 2).
    The authentication mechanism should preserve data throughput without requiring additional payload space.
    \ac{acric} uses the existing \ac{crc} field, preserving payload size and data throughput.

    \item \textbf{Ensure Real-Time Response} (Challenge 4).
    The solution must support the real-time requirements typical of industrial systems.
    \ac{acric} minimizes computational overhead and communication delays by avoiding resource-intensive cryptographic operations, thus ensuring real-time capabilities.

    \item \textbf{Design a Protocol-Agnostic Solution} (Challenge 5).
    The authentication mechanism should apply across various communication protocols; \ac{acric} leverages the widespread presence of the \ac{crc} field, making it adaptable to a broad range of legacy systems.
\end{itemize}

By meeting these security and operational goals, \ac{acric} effectively addresses the identified challenges, offering a practical and robust authentication solution explicitly tailored to legacy industrial communication environments.
\section{Security Model}\label{sec:Model}

In this section, we define the context and assumptions according to which we developed the \ac{acric} framework.
Section~\ref{subsec:SystemModel} outlines our system model, while Section~\ref{subsec:ThreatModel} identifies adversaries capabilities and relevant attacks.

\subsection{System Model}\label{subsec:SystemModel}
We developed \ac{acric} specifically to secure communications in legacy systems that involve older, resource-constrained devices with hard real-time demands for which replacing or significantly upgrading hardware is costly and impractical for these systems.
Thus, \ac{acric} prioritizes backward compatibility and real-time guarantees.
Typical examples include \acp{ics} and in-vehicle networks, widely recognized as representative legacy systems \cite{Castellanos-et-al}, \cite{CANSurvey}.
Besides, given their critical role in public safety and essential services, these domains urgently require practical and compatible security solutions.
A widely adopted and standardized mechanism to ensure reliable data transmission in industrial communication protocols is the use of the \ac{crc} field \cite{CRC3}, \cite{CRC2}.
Since \ac{crc} is already present in most legacy industrial protocols, we propose using it to provide security, preserving its original role for integrity check without altering the message format.

A key feature of our system model is the coexistence of secure and non-secure devices within the same network.
In industrial environments, upgrading all devices simultaneously to new security standards is usually impractical due to logistical constraints, high costs, and the necessity for uninterrupted operation \cite{InvehicleNetworks}, \cite{RetrofittingProtection}.
By allowing for phased security upgrades, we ensure the continuous functionality of critical infrastructure.
To maintain broad applicability, our model abstracts from specific network layouts and emphasizes essential functionalities.
Consequently, it supports various communication protocols, providing authentication mechanisms suitable for both \ac{p2p} and broadcast communication models (we treat multicast and unicast as special cases of these).
Examples of widely adopted protocols compatible with our approach include Modbus~\cite{modbus-application}, CAN-bus~\cite{CAN}, Profibus~\cite{Profibus}, and DNP3~\cite{DNP3}.
This flexibility ensures that \ac{acric} can be effectively applied across different network configurations, enhancing security while maintaining compatibility with existing infrastructures.

\subsection{Threat Model}\label{subsec:ThreatModel}
Our threat model builds on well-established frameworks in previous research (\cite{Bernieri-et-al}, \cite{Katulic-et-al}, \cite{CybersecCAN}, \cite{CANSurvey}), and follows the widely accepted \textit{Dolev-Yao} model~\cite{DV_model}.
Our model defines attacker capabilities and constraints consistently recognized in both networked and cyber-physical systems, ensuring alignment with standard assumptions.
Specifically, we assume an attacker with complete access to network traffic but incapable of compromising hardware or internal cryptographic mechanisms.
Following Kerckhoff's principle~\cite{Kerckhoff_principle}, the attacker is fully knowledgeable about the authentication protocol, relying solely on keys secrecy for protection.
Under these assumptions, potential attacker actions and consequences include:
\begin{itemize}
    \item \textit{Spoofing}: The attacker sends messages appearing to originate from legitimate devices, leveraging compromised or guessed authentication parameters.
    This can result in the execution of unauthorized commands, leading to disruptions in operations and potential safety hazards.

    \item \textit{Tampering}: The attacker intercepts and modifies data in transit without altering the authentication tag, possibly exploiting weaknesses in protocol implementation.
    Such modifications can cause control systems to malfunction or make incorrect operational decisions.

    \item \textit{Replay Attacks}: By capturing legitimate messages and retransmitting them later, the attacker triggers unintended repeated actions.
    This can lead to undesired redundancy, resource exhaustion, or unauthorized activation of system processes.

    \item \textit{Brute-force Attacks}: The attacker systematically tests authentication values until finding a valid one.
    Success could grant unauthorized access or control, compromising system integrity.
\end{itemize}

We explicitly exclude threats such as \ac{dos}, fingerprinting, and side-channel attacks, as these require additional defensive strategies, like network monitoring and anomaly detection, which go beyond cryptographic authentication alone~\cite{CANSurvey}.
\section{Authenticated Cyclic Redundancy Integrity Check}\label{sec:ACRIC}
In Section \ref{subsec:Overview} we first give an overview of \ac{acric}'s principles and functioning, while in Section \ref{subsec:TechnicalOperation} we present its technical details.
Then, in Section \ref{sec:ManageCrypto} we present recommendations for managing the cryptographic material.

\subsection{\ac{acric} Overview}\label{subsec:Overview}
Communication protocols typically use a \acrfull{crc} field to detect transmission errors, calculated using a standardized \acrfull{iv}, a generator polynomial and message content~\cite{CRC}.
This standard \ac{crc} mechanism lacks any security properties, particularly authentication and integrity protection.
Hence, an attacker capable of modifying message payloads can easily recalculate and replace the \ac{crc}, thereby evading detection.
\acrfull{acric} addresses this security gap by repurposing the existing \ac{crc} field to provide both authentication and message integrity.
The core innovation of \ac{crc} lies in the introduction of two complementary enhancements:
\begin{enumerate}
    \item A \textit{secret} \ac{iv}, shared only among authorized entities, is used for \ac{crc} computations, binding message authenticity to the secrecy of this \ac{iv}.
    \item An \textit{OTP encryption}, using single-use secret keys derived from a hash chain, protects the secrecy of the \ac{crc} and, consequently, prevents attackers from deducing the used \ac{iv}.
\end{enumerate}

\begin{figure}[!t]
    \centering
    \includegraphics[scale=0.5]{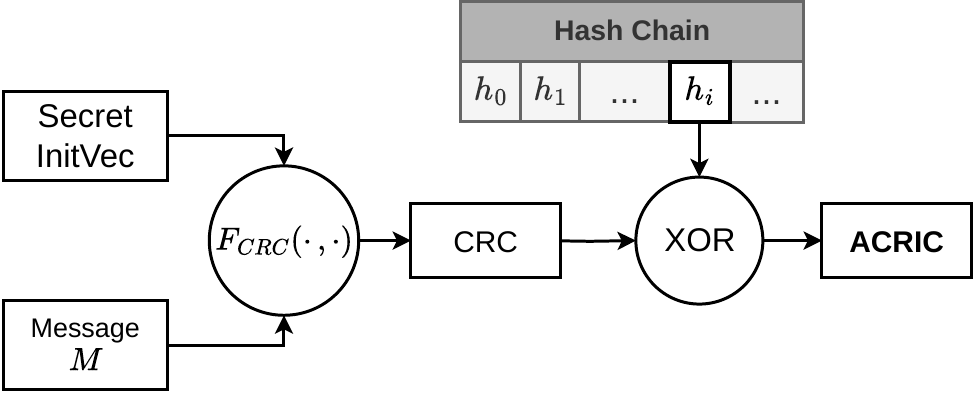}
    \caption{\texttt{ACRIC} computation procedure.
    The secret \ac{iv} and message \(M\) are inputs to the standard \ac{crc} algorithm \(F_{CRC}\left(\cdot, \cdot\right)\).
    The resulting \ac{crc} is \ac{otp} encrypted (XOR-ed) with a hash value to protect the \ac{iv}'s confidentiality.}
    \label{fig:Authentication}
\end{figure}

These enhancements enable \ac{acric} to significantly strengthen legacy communication protocols, ensuring robust authentication and integrity while maintaining full compatibility with legacy systems and interoperability between secured and unsecured devices.
Figure~\ref{fig:Authentication} gives a graphical representation of the \texttt{acric} computation procedure using the secret \ac{iv} and hash values \(h_i\).

\subsection{\ac{acric} Operational Procedure}\label{subsec:TechnicalOperation}
The specific implementation for system initialization and distribution of cryptographic material is out of the scope of this work as it shall be designed according to the specific network topology, constraints, and manufacturer convenience.
Consequently, we assume that upon successful system initialization, each pair of nodes shares the following cryptographic parameters:
\begin{itemize}
    \item \(K_m\): a secret master key;
    \item \(K_s\): a secret session key;
    \item \(H\left(\cdot\right)\): a cryptographic hash function;
    \item \(l\): a secret value for hash chain initialization;
    \item \(\mathcal{H} = \left\{h_0 = H\left(l\right),\, h_i = H\left(h_{i-1}\right),\; i = 1, \dots, N\right\}\): a hash chain of \(N\) elements.
\end{itemize}

\textit{\underline{Phase 1: \ac{crc} Calculation with Secret Initialization}.}
\texttt{ACRIC} computation begins by following the standard \ac{crc} calculation procedure, modified to use a secret \ac{iv} instead of a standardized one.
For an \(n\)-byte \ac{crc}, we define the secret \ac{iv} using the session key \(K_s = \left[k^s_0 || k^s_1 || \cdots || k^s_N\right]\), with each \(k^s_i\) as an 8-bit binary sequence: \(\texttt{InitVec} = \left[k^s_0 || \cdots || k^s_n\right]\).
A matching \ac{crc} thus confirms the authenticity and integrity of the message.
However, using only a secret \ac{iv} is insufficient for security, as attackers could infer it through an exhaustive analysis of message payloads and \ac{crc} values.

\textit{\underline{Phase 2: \ac{otp} Encryption of \ac{crc}}.}
To overcome the limitation described above, \ac{acric} secures the computed \ac{crc} through \ac{otp} encryption.
Specifically, the computed \ac{crc} value is XOR-encrypted using a single-use secret key derived from the hash chain \(\mathcal{H}\):
\begin{equation}\label{eq:authentication}
    \texttt{ACRIC}_i = CRC_i \oplus h_i = F_{CRC}(\texttt{InitVec}, M_i) \oplus h_i,
\end{equation}
where \(F_{CRC}(\cdot, \cdot)\) is the standard \ac{crc} calculation function, \(M_i\) is the \(i\)-th message, and \(h_i\) is the \(i\)-th hash value in the chain.
The hash chain \(\mathcal{H}\) provides a series of single-use encryption keys, ensuring the \ac{otp}'s perfect secrecy property and thus fully protecting \ac{iv} confidentiality \cite{OTP}.

\subsection{Managing the Cryptographic Material}\label{sec:ManageCrypto}
Effective management of cryptographic material is essential when securing legacy industrial systems, especially given their constraints regarding memory and computational resources.
The management of cryptographic material directly impacts the effectiveness of the security measure, the complexity of the deployment, and the operational efficiency.
In the following, we outline practical considerations and offer clear recommendations for managing cryptographic parameters, focusing on system initialization procedures, key management strategies, and hash chain handling.
Table \ref{tab:CryptoManagement} summarizes our discussion.\\
\indent
\textbf{System Initialization.}
Minimizing manual intervention during deployment and maintenance is crucial in large-scale industrial environments due to the inherent complexity and high risk of human error.
Although local initialization - where session keys are derived locally without device interactions - reduces runtime overhead, it demands precise manual configuration of cryptographic parameters.
Such manual setups significantly increase the likelihood of errors, especially in large-scale environments, potentially compromising network security and causing initialization failures.
To simplify and secure the system setup, we then recommend using an interactive initialization strategy.
Specifically, storing only a master key \(K_m\) shared across all network nodes allows devices to dynamically generate session keys, simplifying the system setup although introducing some overhead.
Employing an authenticated Diffie-Hellman key exchange procedure is particularly beneficial, as it provides a secure, flexible, and scalable solution applicable to both \ac{p2p} and broadcast scenarios.\\
\indent
\textbf{Key Management.}
The choice of the key management strategy significantly influences the security and operational efficiency of \ac{acric} in different communication models.
A \textit{Pair-based Key Management} is ideal for \ac{p2p} communications, assigning unique session keys to each pair of communicating devices.
Compared to device-based management, this method simplifies key distribution, parameters synchronization, and reduces storage overhead.
Similarly, a \textit{Group-based Key Management} is recommended for broadcast communication scenarios.
In this approach, devices within logical groups share common cryptographic keys.
Compared to device-based key management - which incurs in high storage overhead and security risks - the group-based method substantially reduces memory requirements and simplifies synchronization.
This approach is particularly advantageous in large systems, as exemplified by its extensive consideration as a feasible key management strategy to introduce authentication within the CAN-bus protocol~\cite{CANSurvey}.
However, a careful implementation of the group-key method is crucial to avoid security vulnerabilities.
Group memberships must be strictly managed, assigning only the minimum necessary permissions aligned to their roles~\cite{CANSurvey}.
Such rigorous management effectively minimizes unauthorized access and mitigates masquerade attack risks.\\
\indent
\textbf{Hash Chain Management.}
The practical and efficient handling of hash chains, which supply the \ac{otp} encryption keys, involves three core components: consumption, storage, and computation.
Regarding \textit{Hash Chain Consumption}, we recommend a forward-consumption strategy, sequentially using hash values from root to tail.
This approach minimizes memory demands compared to a backward-consumption strategy as it does not require pre-computing the entire hash chain.
We highlight that it requires embedding a secret key in the hash function to maintain forward secrecy.
For \textit{Hash Chain Storage}, as well as for key management, we advise for a pair-based (for P2P) or group-based (for broadcast) solution over entity-based storage.
These methods simplify synchronization and significantly reduce memory usage, aligning with requirements of resource-constrained environments and legacy devices.
Finally, for \textit{Hash Chain Computation} we suggest employing a hybrid computation approach, \textit{i.e.}, precomputing part of the hash chain offline and generating the remainder dynamically,
This allows to effectively balance between computational load and memory usage, ensuring both security and operational efficiency.\\
\indent
\textbf{Practical Considerations for Hash Chain Updates.}
Another practical aspect to consider is whether the hash value should be updated after every message transmission or upon successful message reception.
Industrial protocols typically interpret rejected messages as transmission failures, prompting retransmissions a limited number of times before suspending the transmitting device if failures persist \cite{CAN}, \cite{modbus-application}.
In such scenarios, reusing the same hash value for retransmissions generally poses minimal risk.
However, attackers may change the \texttt{ACRIC} value with each retransmission, thus reducing the pool of valid authentication candidates.
While device suspension after repeated transmission failures can deter brute-force attacks, updating the hash value for every retransmission significantly enhances security in critical applications where even a single malicious message could be highly detrimental.
This prevents attackers from systematically narrowing down the set of potential valid authentication tags.
Nonetheless, this solution accelerates hash chain consumption and increases computational overhead due to frequent hash chain refreshes, demanding careful consideration aligned with the application's criticality and specific operational constraints.

\begin{table}[!t]
\renewcommand{\arraystretch}{1.25}
\centering
\caption{Summary of recommended cryptographic material management strategies, highlighting their benefits and potential drawbacks.}
\label{tab:CryptoManagement}
\resizebox{\textwidth}{!}{%
\begin{tabular}{|l|c|c|c|}
\hline
\rowcolor[HTML]{C0C0C0} &
  \multicolumn{1}{c|}{\cellcolor[HTML]{C0C0C0}\textbf{\begin{tabular}[c]{@{}c@{}}Recommended\\Approach\end{tabular}}} &
  \multicolumn{1}{c|}{\cellcolor[HTML]{C0C0C0}\textbf{Benefits}} &
  \multicolumn{1}{c|}{\cellcolor[HTML]{C0C0C0}\textbf{\begin{tabular}[c]{@{}c@{}}Potential\\Drawbacks\end{tabular}}} \\
  
  \hline

    \textit{System Initialization} & \multicolumn{1}{c|}{Interactive Initialization} & \begin{tabular}[c]{@{}c@{}}Minimized human intervention,\\ Simplified system setup\end{tabular} & Increased runtime overhead \\
    
    \hline
 &
  Master key: common key \(K_m\) & \begin{tabular}[c]{@{}c@{}}Reduced storage overhead,\\Simplified system setup\end{tabular} & Single point of failure
   \\
   \cline{2-4} 
    \multirow{-2}{*}{\textit{Key Management}} & Session keys: Pair/Group - based & \begin{tabular}[c]{@{}c@{}}Reduced storage overhead,\\ Simplified synchornization\end{tabular} & \begin{tabular}[c]{@{}c@{}}Group-based approach\\requires careful management\end{tabular} \\
    
    \hline
 &
  Consumption: Forward & Reduced storage overhead & Increased runtime overhead \\ \cline{2-4} 
 &
  Storage: Pair/Group - based & \begin{tabular}[c]{@{}l@{}}Simplified synchronization,\\Reduced storage overhead\end{tabular} & \begin{tabular}[c]{@{}c@{}}Group-based approach\\requires careful management\end{tabular} \\
  \cline{2-4} 
    \multirow{-3}{*}{\begin{tabular}[c]{@{}l@{}}\textit{Hash Chain}\\ \textit{Management}\end{tabular}} & Computation: Hybrid & \begin{tabular}[c]{@{}c@{}}Balances runtime of\\cryptographic computations\end{tabular} & Moderate implementation complexity
   \\
   
   \hline
\end{tabular}%
}
\end{table}
\section{\ac{acric} Security and Performance Evaluation}\label{sec:Evaluation}
In this section, we first assess \ac{acric}'s security guarantees against the security goals and threat model in previous sections.
Subsequently, we evaluate \ac{acric}'s performance regarding computational and transmission overhead.
We provide the source codes for experimental evaluations in a GitHub repository\footnote{\url{https://github.com/aleLtt/ACRIC}}.

\subsection{Security Assessment}\label{subsec:Security}
To assess \ac{acric}'s security properties, we assume that the cryptographic parameters \(K_m\), \(K_s\), \(l\), and \(\mathcal{H}\) remain confidential between the legitimate parties.
We then address each security goal presented in Section~\ref{sec:SecGoals} separately.

\textbf{Authentication}.
Given a fixed message \(M\), we can model Equation \eqref{eq:authentication} as:
\begin{equation}\label{eq:Model}
    C = F(A) \oplus B,
\end{equation}
where we abstract the message \(M\) from the input to the \ac{crc} function.
Consider the legitimate tuple \(\left(A_1, B_1, C_1\right)\) with each value being \(n\) bits long, and define the number of possible inputs \(A\) producing the same output of \(A_1\) as: \(m = \left|\left\{A \; s.t. \; F\left(A\right) = F\left(A_1\right)\right\}\right|\).
To assess \ac{acric}'s authentication guarantees, we evaluate the \texttt{ACRIC} collision probability \(P_c\) when authenticating a message with incorrect parameters \(\left(A_2, B_2\right)\).
We identify three possible scenarios.
\begin{enumerate}
    \item \underline{\(A_2 = A_1 \,,\, B_2 \neq B_1\)}: Having a collision \(C_2 = F\left(A_1\right) \oplus B_2 = F\left(A_1\right) \oplus B_1 = C_1\) implies \(B_2 = B_1\).
    However, \(B_2 \neq B_1\) by assumptions, thus:
    \begin{equation}
        P_1 = Pr\left[C_2 = C_1 | B_2 \neq B_1\right] = 0.
    \end{equation}

    \item \underline{{\(A_2 \neq A_1 \,,\, B_2 = B_1\)}}: Having a collision \(C_2 = F\left(A_2\right) \oplus B_1 = F\left(A_1\right) \oplus B_1 = C_1\) implies \(F\left(A_2\right) = F\left(A_1\right)\).
    This means that there must be multiple values \(A\) (corresponding to \ac{iv}) that, for a given message, produce the same \ac{crc} value.
    By definition, excluding \(A_1\), there are \(m-1\) such values.
    On the other hand, there are \(2^n -1\) possible values \(A \neq A_1\).
    Therefore:
    \begin{equation}
        P_2 = Pr\left[C_2 = C_1 | A_2 \neq A_1\right] = \frac{m-1}{2^n - 1}
    \end{equation}

    \item \underline{{\(A_2 \neq A_1 \,,\, B_2 \neq B_1\)}}: Having a collision \(C_2 = F\left(A_2\right) \oplus B_2 = F\left(A_1\right) \oplus B_1 = C_1\) implies \(B_2 = F\left(A_2\right) \oplus F\left(A_1\right) \oplus B_1\).
    Consequently:    
    \begin{equation}\label{eq:P3}
        \forall A_2, \; \exists ! \; B_2^* \;s.t.\; C_2 = C_1 \Rightarrow
        \begin{cases}
            \forall A_2 \neq A_1 \;s.t.\; F\left(A_2\right) \neq F\left(A_1\right) \;, \; \exists ! \; B_2 \neq B_1 \\
            \forall A_2 \neq A_1 \;s.t.\; F\left(A_2\right) = F\left(A_1\right) \;, \; \nexists \; B_2 \neq B_1.
        \end{cases}
    \end{equation}
    Thus, among all values, exactly \(\left(m-1\right)\) yield to \(F\left(A_2\right) = F\left(A_1\right)\), giving 0 ways to pick \(B_2 \neq B_1\) to collide; the remaining \(\left(2^n-1\right)-\left(m-1\right) = 2^n-m\) values of \(A_2\) have \(F\left(A_2\right) \neq F\left(A_1\right)\), each providing exactly one suitable \(B_2 \neq B_1\).
    Therefore, \(P_3 = Pr\left[C_2 = C_1 | A_2 \neq A_1, B_2 \neq B_1\right] = \frac{2^n - m}{\left(2^n-1\right)^2}\).  
\end{enumerate}

In conclusion, \texttt{ACRIC} collision probability \(P_c\) is:
\begin{equation}\label{eq:CollisionProb}
    P_c = P_1 + P_2 + P_3 = \frac{m-1}{2^n - 1} + \frac{2^n - m}{\left(2^n-1\right)^2}.
\end{equation}
\begin{definition}[Injectivity in the Initialization Vector]
A Cyclic Redundancy Check (CRC) function \( F(\cdot, \cdot) \) is said to be \textbf{injective in the initialization vector} if, for every fixed message \( M \), distinct initialization vectors produce distinct CRC values. Formally: \(\forall M, \, I \neq I' \implies F(I, M) \neq F(I', M)\).
\end{definition}

In particular, when \(F_{CRC}\left(\cdot,\cdot\right)\) is injective in the \ac{iv}, meaning that for a given message there exists one single \ac{iv} value that produces the target \ac{crc} value, we have \(m = 1\).
Hence, Equation~\ref{eq:CollisionProb} simplifies to: \(P_c \sim \frac{1}{2^n}\), matching the intuitive security bound of randomly guessing an \(n\)-bit authentication tag.
We experimentally evaluated the injectivity property in the \ac{iv} of 41 different \ac{crc} functions provided by the \texttt{crcmod}~\footnote{\url{https://crcmod.sourceforge.net/}} Python library, showing that \textit{all the considered \ac{crc} functions are injective in the initialization vector}.

Beyond collisions, we also analyzed the potential exploitation of \ac{crc} linearities.
If \ac{crc} functions were linear, an attacker could craft a valid message \(M' = M \oplus X\) without knowing the security context: \(\texttt{ACRIC}' = \texttt{ACRIC} \oplus CRC(X) = (CRC(M) \oplus h) \oplus CRC(X) = CRC(M \oplus X) \oplus h = CRC(M') \oplus h\).
However, even under this assumption, the attacker still needs to first learn the \ac{iv} used, since we showed that a \ac{iv} collision is not possible.
In addition, we also evaluated the linearity of the previously considered functions.
Our evaluation shows that in their standard version, functions with \(\ac{iv}= 0\) and \(\texttt{XOR-out} = 0\) are linear; however, when using a different \ac{iv}, as we do in \ac{acric}, all these functions lose linearity, thus ensuring \ac{acric}'s resistance against such attacks.

\textbf{Resistance to Brute-Force Attacks}. \ac{acric}'s primary limitation is the finite length of the \ac{crc} field, typically ranging from 8 to 64 bits, potentially making it vulnerable to brute-force attacks.
However, a common and recommended practice in \ac{ot} and \ac{ics} environments is to limit the number of allowed failures to a few messages, after which devices are suspended and alerts are triggered \cite{CAN}, \cite{modbus-application}, \cite{NIST-Security}.
This widely adopted procedure effectively prevents brute-force attacks, thus providing robust protection for \ac{acric}.
In addition, as discussed in Section~\ref{sec:ManageCrypto}, by changing the hash value used for \ac{otp} encryption for every message, we completely prevent brute-force attacks, reducing the attack success probability to the same as a random guess.
However, this approach consumes the hash chain faster, requiring updates to the chain more frequently.

\textbf{Integrity Protection}. When a verifier receives message \(M'\) with an \texttt{ACRIC} value computed using message \(M\), it applies the correct \ac{iv} and hash values to verify authenticity.
Hence, evaluating the integrity protection property involves assessing the probability that two distinct messages, \(M\) and \(M'\), generate the same \ac{crc} value.
Since \ac{crc} functions are not true random functions, the probability may strongly depend on the specific function.
We experimentally evaluated the aforementioned probability for several functions in the \texttt{crcmod} library.
The results show that the probability of such collisions is effectively negligible, thus ensuring strong integrity protection.

\textbf{Replay Attack Resistance}. Messages replayed within the same session will have the correct \ac{iv} but an incorrect hash value, corresponding to authentication Case 1, resulting in zero probability of successful replay.
Conversely, messages replayed across different sessions align with authentication Cases 2 or 3.
Given that the \ac{crc} functions are injective with respect to different \ac{iv}, the success probability for such replay attacks is \(2^{-n}\), equivalent to a random guess.
This demonstrates \ac{acric}'s robust resistance against replay attacks.\\
\indent
In conclusion, this security analysis confirms that \ac{acric} meets all the specified security objectives and effectively resists the identified attacks.
Nonetheless, we emphasize that the security of \ac{acric} fundamentally depends on preserving the secrecy of both the \ac{iv} and the hash chain \(\mathcal{H}\).
Any compromise of these components would critically undermine \ac{acric}'s security, highlighting the importance of rigorously protecting these elements.

\subsection{Performance Evaluation}\label{subsec:Performance}
We assessed \ac{acric} integration in mixed systems, with secure and non-secure environments, measured its computational overhead, and compared its performance with traditional \ac{hmac}-based authentication methods.

\begin{figure}[!h]
    \centering
    \includegraphics[width=0.65\linewidth]{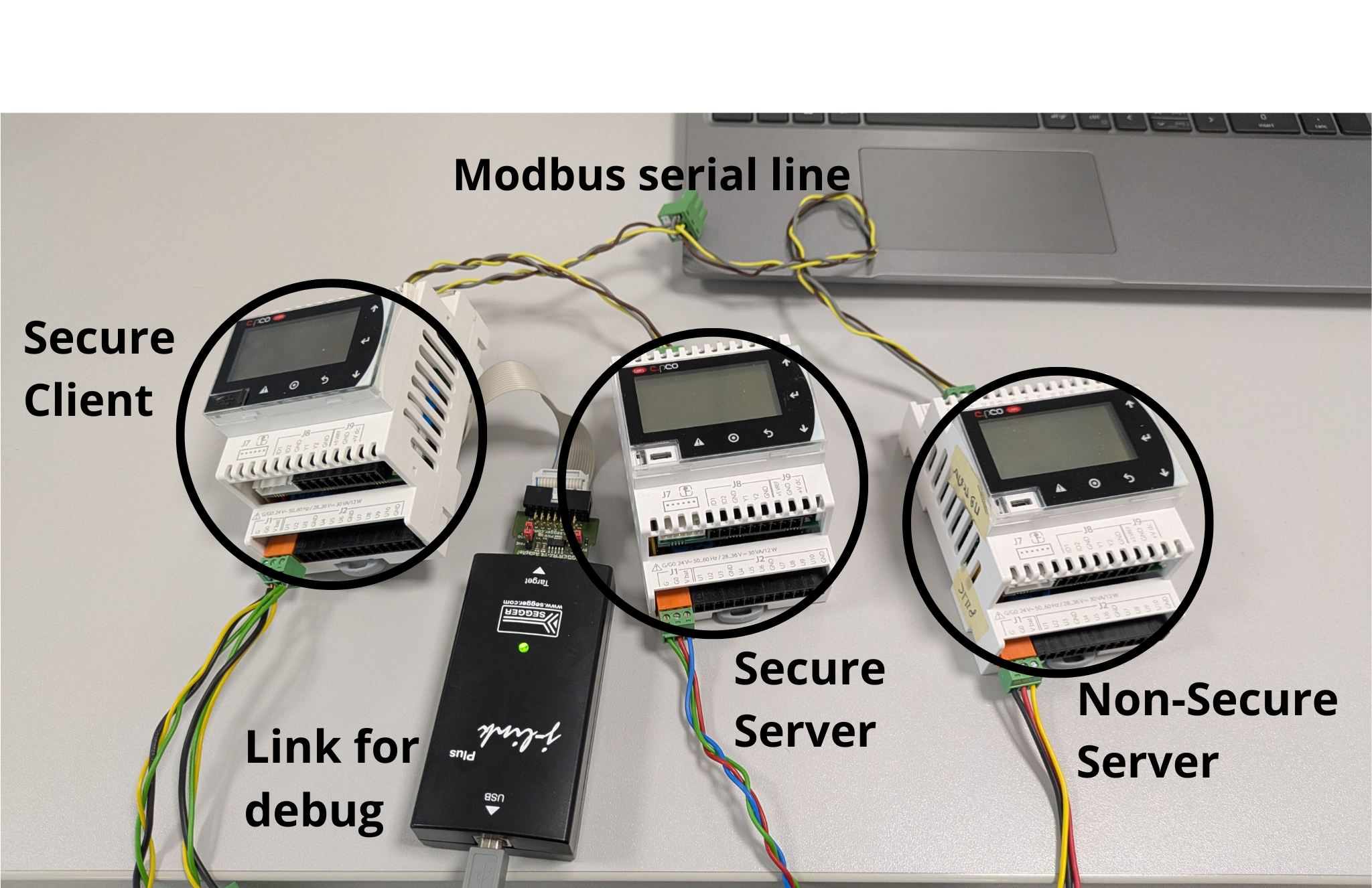}
    \caption{Testbed with industrial devices for \ac{acric} validation and evaluation.}
    \label{fig:Setup}
\end{figure}

\textbf{Experimental Setup.}
Our experimental setup integrates \ac{acric} into the Modbus protocol, a widely used industrial standard prevalent in legacy and resource-constrained environments.
Figure~\ref{fig:Setup} illustrates our testbed, comprising three commercially available industrial devices from CAREL, each equipped with an STM32F4xx microcontroller and 16 MB external DRAM.
ACRIC was implemented using proprietary software with cryptographic operations supported by the emCrypt library\footnote{\url{https://www.segger.com/products/security-iot/emcrypt/}}, ensuring accurate and realistic performance measurements representative of industrial conditions.
For the management of cryptographic material, we opted for a pair-based forward-consumption hash-chain management strategy with hybrid computation.

\begin{figure}[!h]
    \centering
    \includegraphics[width=0.7\linewidth]{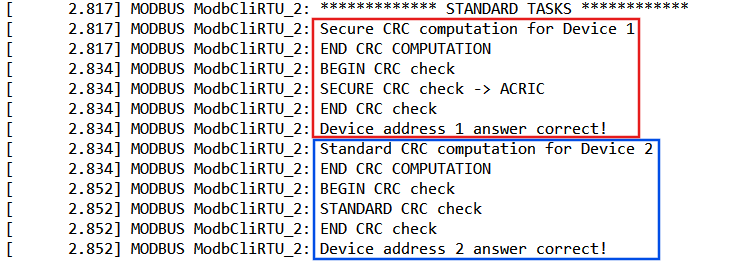}
    \caption{Debug prompt showing \ac{acric} enabled interoperability between secured (blue square) and non-secure (red square) devices.}
    \label{fig:interoperability}
\end{figure}

\textbf{Interoperability.}
We assessed \ac{acric}'s interoperability by conducting tests within a testbed that included both secured and non-secured Modbus devices.
The seamless and stable communication observed between secured and non-secured devices confirmed \ac{acric}'s effective backward compatibility, highlighting its suitability for integration into existing industrial systems and allowing for phased system upgrades (Figure \ref{fig:interoperability}).


\textbf{Computational Overhead.}
We compared the execution times of standard Modbus messages with those secured by \ac{acric}.
Our results show an average overhead of approximately \(4 \mu s\) per message, demonstrating minimal latency impact suitable for real-time industrial operations.
For comparative purposes, we also implemented SHA-1 \ac{hmac} authentication, which exhibited a significantly higher overhead of approximately \(371 \mu s\) per message.
This direct comparison clearly underscores \ac{acric}’s efficiency and suitability over traditional methods for resource-constrained industrial applications.
These results confirm that \ac{acric} meets all operational goals identified in Section~\ref{sec:OperationalGoals}, providing robust security while maintaining essential industrial requirements of data throughput, backward compatibility and minimal computational overhead, making it an ideal solution for retrofitting security in legacy systems.

\section{Conclusions}\label{sec:Conclusions}

In this paper, we addressed the problem of retrofitting security within legacy industrial and \ac{ot} systems, identifying practical challenges hindering the real-world adoption of authentication solutions proposed in the literature.
These include enabling interoperability between secure and non-secure devices, preserving message format and data throughput, minimizing deployment costs and overhead, and ensuring compatibility with mixed systems.
To overcome all these limitations at once, we designed \acrfull{acric}, an authentication mechanism that leverages cryptographic computation of the existing \acrfull{crc} field, ensuring message authentication and integrity protection.
Compared to existing solutions, \ac{acric} preserves the original frame structure and does not require additional hardware, ensuring full backward compatibility and throughput for legacy protocols.
Moreover, its protocol-agnostic design enables seamless integration into diverse industrial scenarios, supporting both point-to-point and broadcast communications.
Formal analysis and experimental validation with commercial devices demonstrate \ac{acric}’s robust security and minimal computational overhead (\(\sim 4 \mu s\)), confirming its practicality for securing legacy industrial applications.


%
%
%

\bibliographystyle{splncs04}
\bibliography{Bibliography}

\end{document}